\documentclass[journal=jacsat,manuscript=article]{achemso}

\usepackage{chemformula} 
\usepackage[T1]{fontenc} 

\author{Poonam Kumari}
\affiliation{A shared footnote}
\author{Cyrille Barreteau}
\affiliation{SPEC, CEA, CNRS, Universit\'e Paris-Saclay, CEA Saclay, Gif-sur-Yvette F-91191, France}
\author{Alexander Smogunov}
\affiliation{SPEC, CEA, CNRS, Universit\'e Paris-Saclay, CEA Saclay, Gif-sur-Yvette F-91191, France}
\email{alexander.smogunov@cea.fr}

\title[An \textsf{achemso} demo]
  {Modelling spin-orbitronics effects at interfaces and chiral molecules}

\begin{document}







\begin{abstract}
Using orbital angular momentum (OAM) currents in nanoelectronics, for example, for magnetization manipulation via 
spin-orbit torque (SOT), represents a growing field known as "spin-orbitronics". Here, using the density functional theory (DFT) and 
the real-time dynamics of electronic wave packets, we explore a possibility of generation and propagation of orbital currents 
in two representative systems: an oxidized Cu surface (where large OAMs are known to form at the Cu/O interface) and a model molecular junction 
made of two carbon chains connected by a chiral molecule.
In the Cu/O system, the orbital polarization of an incident wave packet from the Cu lead is strongly enhanced at the Cu/O interface but then rapidly decays 
in the bulk Cu due to orbital quenching of asymptotic bulk states. 
Interestingly, if a finite transmission across the oxygen layer is allowed (in a tunnel junction geometry, for example),
a significant spin-polarization of transmitted (or reflected) currents is instead predicted which persists at a much longer distance
and can be further tuned by an applied in-plane voltage.
For the molecular junction, the mixing of the carbon $p_x$ and $p_y$ (degenerate) channels by the chiral molecular orbital gives rise not only to an efficient 
generation of orbital current but also to its long-range propagation along the carbon chain.

\end{abstract}

\section{Introduction}
It is well known that spin–orbit coupling (SOC) provides an efficient route for manipulation of the magnetization via spin–orbit torque (SOTs)\cite{SOT1}. SOT is generated when an electric current passing in nonmagnetic material induces a transverse spin current due to SOC and broken inversion symmetry\cite{AChernyshov2009} which can in turn affect the magnetization of an adjacent ferromagnetic layer. 
In these systems with broken inversion symmetry Spin Hall Effect (SHE) \cite{SHE} (spin moment separation) and Spin Edelstein Effect (SEE) \cite{SEE} (spin moment accumulation at the interface) are also often reported and are employed to generate SOT\cite{SOT2, SOT3}. 

Recently, enhanced SOT-related effects have been observed in materials with relatively low SOC like Ti \cite{ChoiNat2023}, Cu/Ferromagnet \cite{HAn2023}, LaAlO$_3$/SrTiO$_3$  \cite{EHamdiNat2023}, Ni/Ti bilayer \cite{HayshiNE2024}, opening up an avenue for investigation at a more fundamental level. It has been argued that another fundamental degree of freedom of an electron, the Orbital Angular Momentum (OAM), may play a crucial role in these materials, either directly, by providing orbital currents, or indirectly, when orbital currents are converted to the spin ones. 
The prominent OAM texture in $k_\parallel$-space (parallel to the interface) can 
be generated by the lack of inversion symmetry accompanied with a strong 
interfacial potential gradient which can mix specific atomic orbitals. 
This effect called Orbital Rashba Effect (ORE) is similar to its spin counterpart but do not require SOC.\cite{ORE} In the presence of this chiral OAM texture, a shift of the occupation function by an external electric field will result in a finite OAM accumulation 
giving rise to the Orbital Edelstein Effect (OEE)\cite{OEE}, similarly to the SEE, which can be however much more significant.

In this letter, based on density functional theory calculations, we addressed two main questions.
Firstly, we wanted to see if the OAM texture in $k_\parallel$-space, generated by the ORE at the interface, can result in a finite transverse orbital current when electrons are reflected out of the interface, as it is shown schematically in Figure~\ref{19copper_o}a. 
Electrons are assumed to be injected at the interface by a small in-plane voltage shifting occupations of electronic states in the $k_\parallel$-space. 
We were, in particularly, motivated by the recent work of D. Go and co-workers\cite{DGoPRB2021} who reported a significant OAM at the oxidized Cu surface due to strong hybridization between surface Cu $d$ states and O $p$ states and the formation of hybrid surface states 
right at the Fermi energy. 
Our results suggest, however, that an electron wave packet loses rapidly its orbital moment when moving away from the oxidized surface. 
However, in the tunnel junction regime, the interplay between this short-range orbital current and the spin degree of freedom generates a noticeable spin-polarization of both transmitted and reflected currents, even though the spin-orbit coupling is rather weak in Cu.   
We argue that the degree of this spin-polarization can be tuned by the applied in-plane voltage. 
This result can be related to the effect known as  
chirality-induced spin selectivity (CISS) consisting in a partial spin-polarization 
of electron current when electrons travel across a chiral structure.  
Second question we address is the possibility of long-range propagation of orbital moment which did not seem to be the case for the Cu/O system.
By inspecting the symmetry of electronic bands, we choose a simplistic carbon chain where $p_x-$ and $p_y$-symmetry (degenerate) bands are both available at the Fermi energy and can therefore sustain a finite orbital current. To make a mixing of these bands by breaking the axial symmetry of the system we introduce a chiral molecule, called 1,5-diamino-[4]cumulene \cite{MGarnerCS2019}, in the junction. Here, as a proof of principle, we successfully demonstrate the efficient generation of the long-range propagating orbital current at the Fermi energy with the chiral molecule acting therefore as a perfect OAM filter, 
the effect which can be referred as the chirality-induced orbital selectivity (CIOS) in analogy with its spin counterpart, CISS.

\section{Results and discussion}
\begin{figure}
\includegraphics[width=16cm]{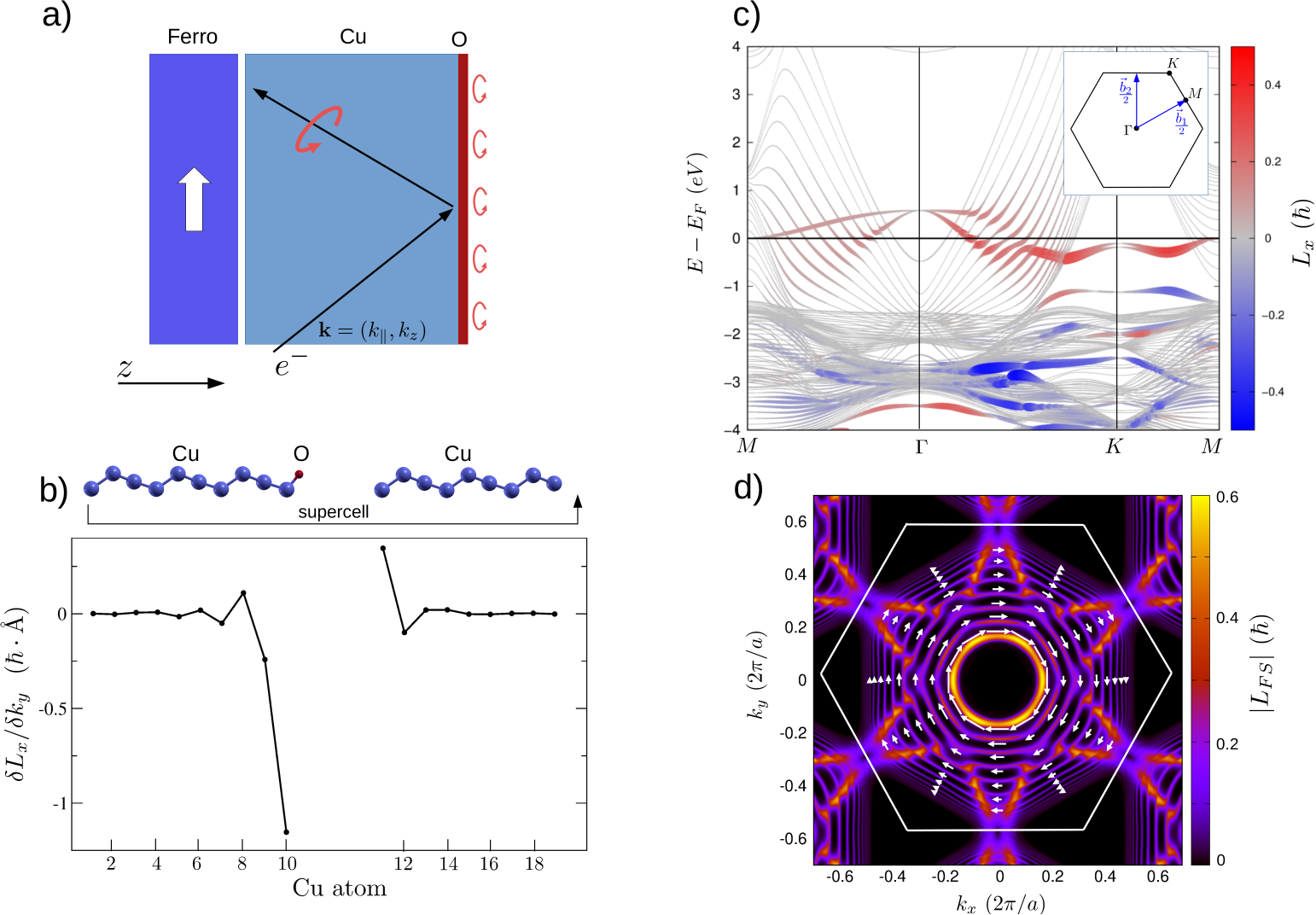}
\caption{\label{19copper_o}
DFT results for the oxidized Cu surface showing the formation of strong orbital moments at the interface: 
a) schematic illustration of possible orbital moment polarization of electron wave packets after reflection 
out of the Cu/O surface;
b) orbital moment (per Cu atom) accumulation for the 19-layer oxidized Cu slab upon application of an electric field.
Electric field in the $y$ direction induces the orbital moment along the $x$ axis;
c) band structure of the Cu/O slab along specific $k_\parallel$-path. The color and the size of circles 
denote the amount of $L_x$ moment; special k-points as well as reciprocal lattice vectors, $\vec b_1,\vec b_2$, 
are shown on the inset;
d) OAM  texture in the 2D Brillouin zone at the Fermi energy; 
the magnitude and the direction of OAM are encoded in the color and shown by arrows (of varying length).
}
\end{figure}

We start with an oxidized copper surface which we model
by a Cu(111) slab of 19 layers with a monolayer of oxygen deposited on one of the Cu surfaces, which is similar to [\hspace*{-5px}\citenum{DGoPRB2021}].
The supercell, set up for subsequent transport calculations, is shown in Figure~\ref{19copper_o}b. It is made of the left part of 10 Cu atoms (and adsorbed Oxygen)
and the right part of 9 Cu atoms separated by 9.2~{\AA}~ in order to avoid interactions between them.
This supercell is periodically repeated in all the three directions.
The Cu layer attached to the O layer is termed as Cu$_{10}$, its $d$ orbitals turn out to hybridize strongly with $p$ orbitals of Oxygen which results in the formation of hybrid Cu/O states around the Fermi energy.\cite{DGoPRB2021} 
We present in Figure~\ref{19copper_o} different OAM quantities which have been calculated according to formulas similar to [\hspace*{-5px}\citenum{DGoPRB2021}] (see also Supporting information).  
We can calculate, for example, the accumulation (per atom) of the OAM due to an applied external in-plane electric field.
Figure~\ref{19copper_o}b shows the $x$ component of such induced OAM per Cu layer due to the field in the $y$ direction
normalized by the wave vector displacement $\delta k$ of the Fermi surface due to the field (Supporting information, Eq.~(17)). 
One observes that the accumulation of the OAM at the surface Cu$_{10}$ is very high but decays very rapidly as one moves away from the surface. We also calculate the bandstructure of the slab along specific $k_\parallel$-path in the 2D Brillouin zone as shown in Figure~\ref{19copper_o}c. 
For each state, the $x$ component of the OAM, projected onto the  Cu$_{10}$ layer, is also provided encoded into the circle radius as well as its color. Finally, the OAM texture in $k_\parallel$-space at the Fermi surface is presented in Figure~\ref{19copper_o}d
(Supporting information, Eq.~(15)).
The magnitude of OAM is shown by the color and the arrows indicate its direction. 
The quantities presented in Figure~\ref{19copper_o} are overall in a pretty good agreement with those of [\hspace*{-5px}\citenum{DGoPRB2021}] and confirm the emergence of hybrid electronic states at the Cu/O interface bearing high OAM.

\begin{figure}[t]
\begin{center}
\includegraphics[width=16cm]{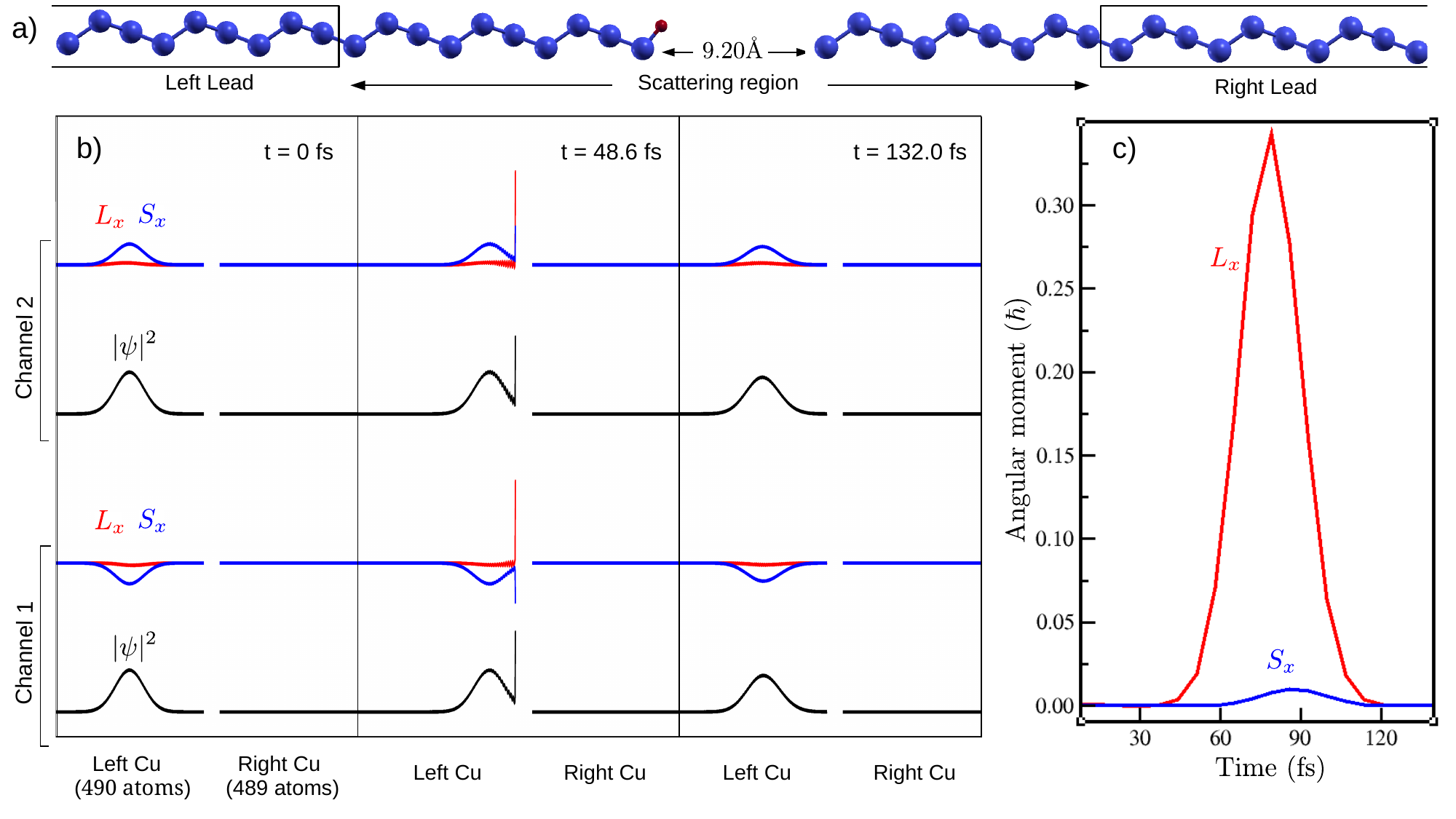}
\caption{\label{wp_surface}
Wave packet propagation from the left Cu lead to the Cu/O surface: 
a) the structure consisting of the left Cu lead, the scattering region and the right Cu lead;
b) snapshots of the wave packet localization function as well as its spin and orbital angular moments (per Cu atom) 
at three time steps;
c) total $L_x$ (in red) and $S_x$ (in blue) summed over two channels.  
}
\end{center}
\end{figure}

As the OAM accumulation decreases quickly out of the Cu/O interface, it has been argued that the orbital current will die down quickly as well.\cite{DGoPRB2021} We wanted to explore this point further and check if the interface OAM may have any effect on the spin current. In order to do this we consider our slab of Figure~\ref{19copper_o}b as a scattering region attached on both sides to semi-infinite Cu leads, as shown in Figure~\ref{wp_surface}a, and carry out transport calculation within the standard Landauer-B\"uttiker approach. 
For a fixed $k_\parallel$ and a chosen energy we construct a Gaussian wave packet in the middle of the left lead and propagate it in time through the scattering region. At the same time we monitor its OAM, $L$, calculated at each atomic site.

We start with $k_\parallel=0.15\vec b_2$ (along the $y$) where a very large OAM (directed along the $x$) 
is found to form at the Cu/O interface at the Fermi energy, as seen in Figure~\ref{19copper_o}d.     
At this k-point two degenerate bands (corresponding to up and down spins) cross the Fermi level in the Cu lead, providing two transport channels (Figure~S1).

Figure~\ref{wp_surface}b shows snapshots of the wave packet propagation from the left lead (left panels) 
to the right one (right panels) with intermediate time when it crosses the scattering region (middle panels).
For each time we show the norm of the wave packet ($|\psi|^2$) as well as 
its orbital ($L_x$) and spin ($S_x$) moments, shown in black, red, and blue, respectively.
Initially, deep inside the Cu lead, $L_x$ is negligible in both channels while $S_x$ is equal in magnitude but opposite in direction, 
resulting in zero overall spin moment (when summed over two channels).
When wave packets reach the surface (middle panels), a sharp increase of $L_x$ is observed 
at surface Cu atoms because of strong Cu/O hybridization.
It is positive for both channels which correlates with the OAM at this $k_\parallel$-point 
calculated for the slab (Figure~\ref{19copper_o}d).
Finally, both wave packets are fully reflected (right panels), $L_x$ dies down quickly 
and no orbital current is observed in neither of two channels, whereas
$S_x$ remains finite but again sums up to zero overall spin moment.
To summarize these results, we summed up the $L_x$ and $S_x$ over two channels 
and also over all the atoms and printed out in Figure~\ref{wp_surface}c
the total values at 20 equally spaced time steps.
One can see that $L_x$, initially close to zero, peaks up around the time wave packets reach the surface Cu atom and then decays and goes to zero as wave packets are reflected back. We also notice that $S_x$ on the other hand remains negligible. 
No orbital or spin currents are therefore generated, unfortunately.

\begin{figure}[t]
\includegraphics[width=16cm]{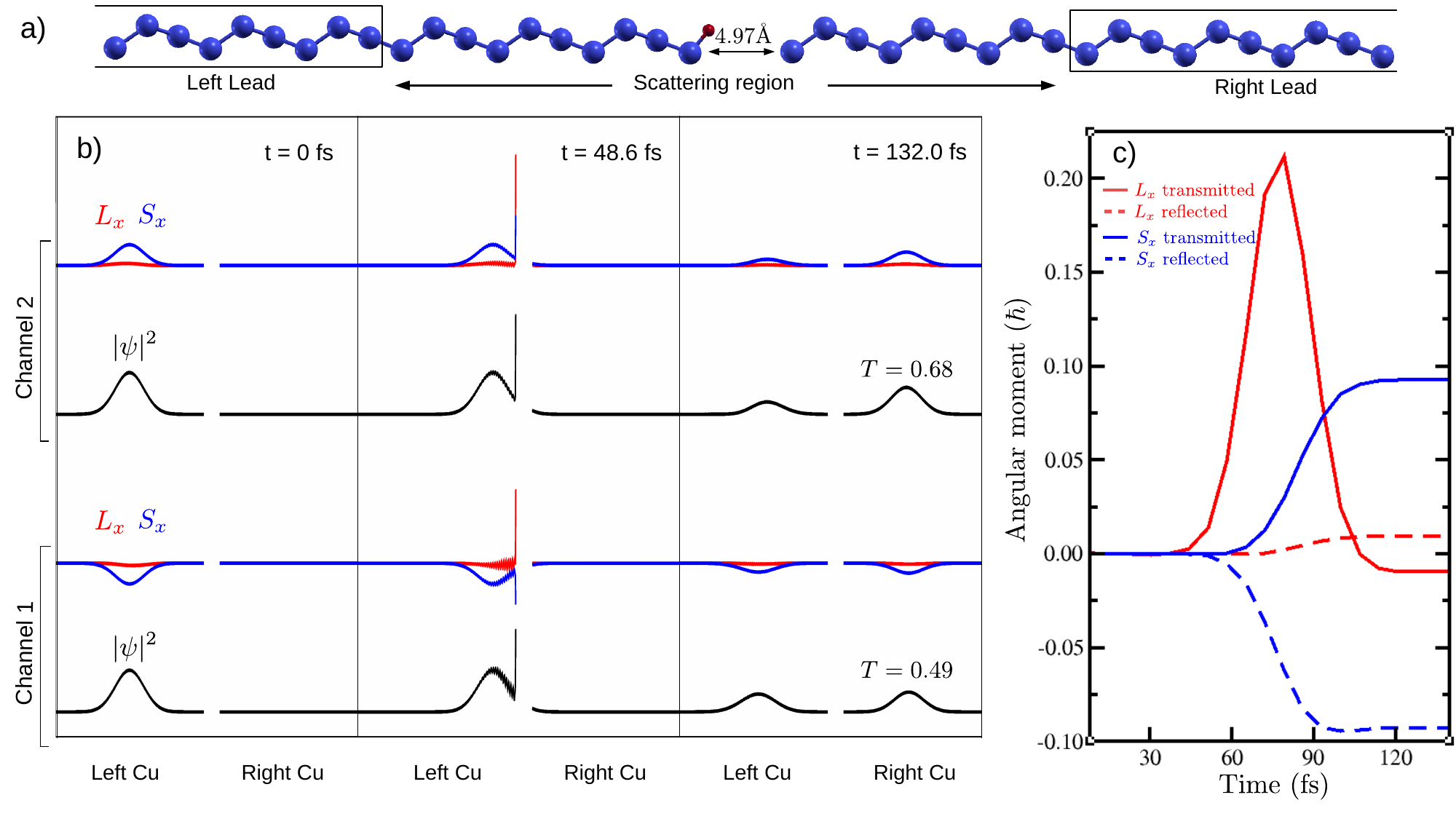}
\caption{\label{wp_junction}
Wave packet propagation in the CuO junction: a) the structure;
b) snapshots of the wave packet at three time steps. Tor the finite time the transmission probability, $T$, is also 
indicated for both channels; 
c) $L_x$ and $S_x$ calculated for transmitted (solid) and reflected (dashed) parts of wave packets and summed over both channels.  
}
\end{figure}

For this structure, due to the large vacuum separation of 9.2{\AA}, wave packets are fully reflected loosing their orbital polarization.
We therefore pose a question of what would happen if a part of a wave packet could propagate across the Cu/O interface.
Hence, we consider a smaller vacuum separation between Cu$_{10}$ and Cu$_{11}$ atoms, 4.97 {\AA}, as shown in Figure~\ref{wp_junction}a. 
Like in the previous case we added two Cu leads to both ends and performed a similar transport analysis, the results
are shown in Figure~\ref{wp_junction}b.
Now as wave packets get transmitted through the scattering region, we see some interesting results. 
In particular, one can notice that the transmitted parts of two wave packets
are not the same: namely, the transmission probability is about 49\% and 68\% for the channel 1 and 2, respectively, as it is
indicated on the right panel in Figure~\ref{wp_junction}b. 
As a consequence, a small but finite amount of $L_x$ is generated in both transmitted (summed over the right 
part) and reflected (summed over the left part) contributions, as presented in Fig.~\ref{wp_junction}c. 
Importantly, this effect of filtering is much more pronounced for the spin moment, $S_x$.
Therefore, the total conductance from this k-point turns out to be spin-polarized of about 16\%. 
This result is important because even though the orbital current itself, generated at the Cu/O interface, 
is very weak, it transforms to quite a significant spin-polarization of the current due to (quite moderate)
SOC in copper. We have further verified that both SOC (needed to activate the spin channel) and the oxygen layer (needed to generate interfacial orbital moments) are important to produce filtering effects discussed above. Indeed, if the oxygen layer is removed or SOC effects are neglected the calculations produce no spin-polarization of the conductance (see Supporting information). Of course, a crucial condition is to have a finite transmissions across the oxygen layer, like in the present case of the Cu/O junction.
The difference in transmissions between the two channels can be attributed to the spin-orbit coupling (SOC) term in the total electronic Hamiltonian, $H_{\rm SOC}=\lambda \hat{\bf L}\cdot \hat {\bf S}$, where $\lambda,~\hat{\bf L}$ and $\hat {\bf S}$ are spin-orbit coupling constant, orbital and spin momentum operators, respectively. If the orbital momentum operator is replaced by its mean value, ${\bf L}$, this term would produce a spin-dependent contribution to the total potential, $\lambda {\bf L} \hat {\bf S}$.
If ${\bf L}$ is large enough that will result in a significant difference of potential barriers across the Cu/O interface for electrons with a spin parallel or anti-parallel to interfacial orbital moment, even if SOC parameter $\lambda$ is small. As a consequence, two spin channels will have slightly different transmissions, as found in our calculations.

\begin{figure}[t]
\includegraphics[width=16cm]{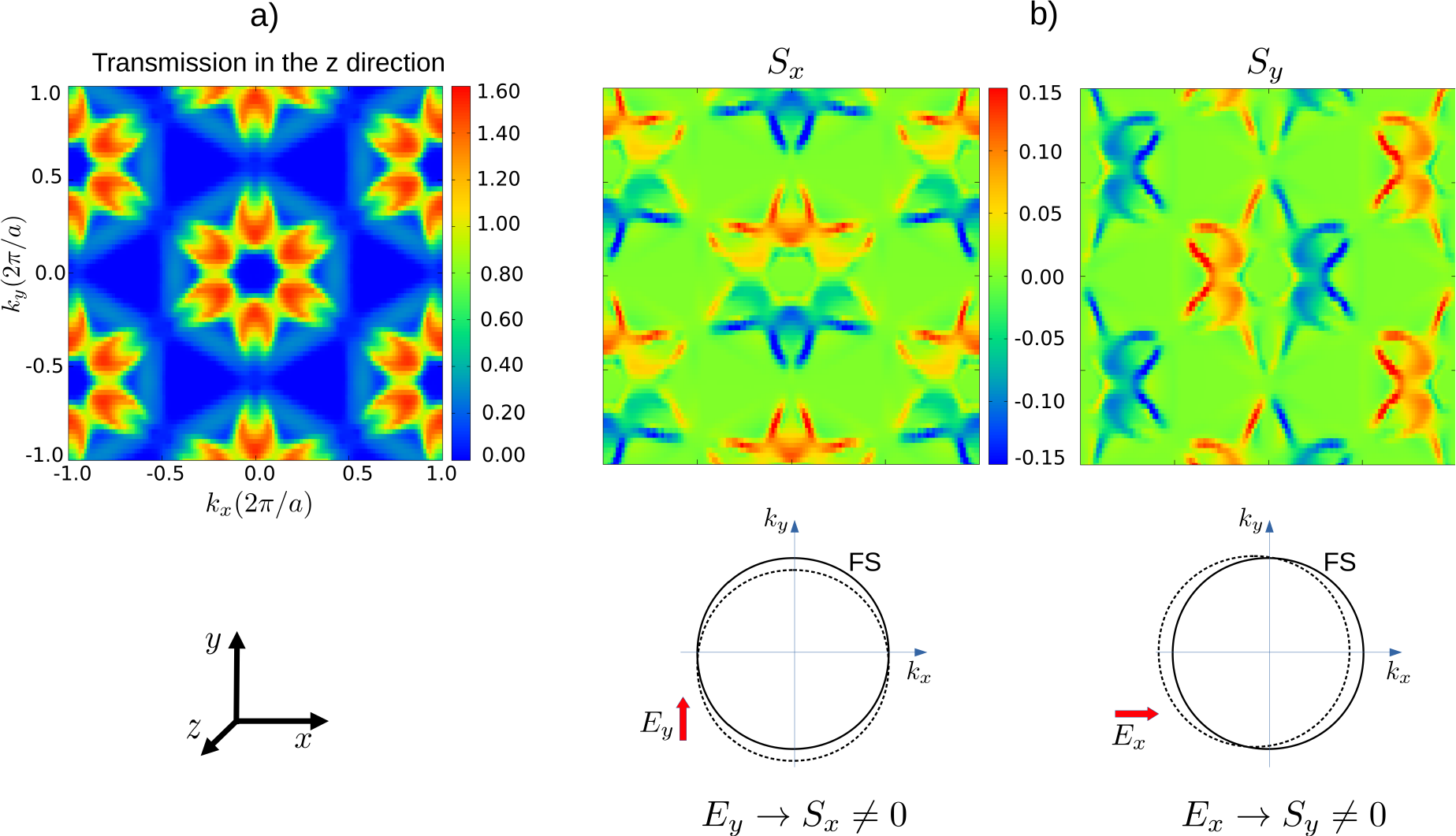}
\caption{\label{2dplots}
2D maps within the $xy$ Brillouin zone for the Cu/O junction of: 
a) the transmission function at the Fermi energy (conductance); 
b) $S_x$ (left) and $S_y$ (right) components of the spin current at the Fermi energy.
The total amount of the filtering can be tuned by an electric field directed along the $y$ or $x$ directions shifting the Fermi surface, FS, as demonstrated schematically on the bottom. 
}
\end{figure}

Inspired by these results, we present in Figure~\ref{2dplots} complete data within the two-dimensional Brillouin zone. 
We show, in particular, the total transmission (left panel) as well as $S_x$ and $S_y$ spin currents (middle and right panels, respectively) calculated at the Fermi energy. One can notice that $S_x$ is antisymmetric with respect to the $x$ axis while $S_y$ -- with respect to the $y$. Therefore, at equilibrium, when all the electronic states up to the Fermi energy are occupied, the total integrated $S_x$ and $S_y$ are zero due to compensation from $k$ and $-k$ states. 
This symmetry can be however perturbed by applying an external in-plane electric field due to corresponding shift of the 2D Fermi surface as shown schematically in Figure~\ref{2dplots}b (bottom). One can see, for example, that the field along the 
$y$ direction should induce a finite $S_x$ and similar for the field in the $x$ direction. 
The induced spin-polarization is orthogonal to the field as it should be.

\begin{figure}[t]
\includegraphics[width=15cm]{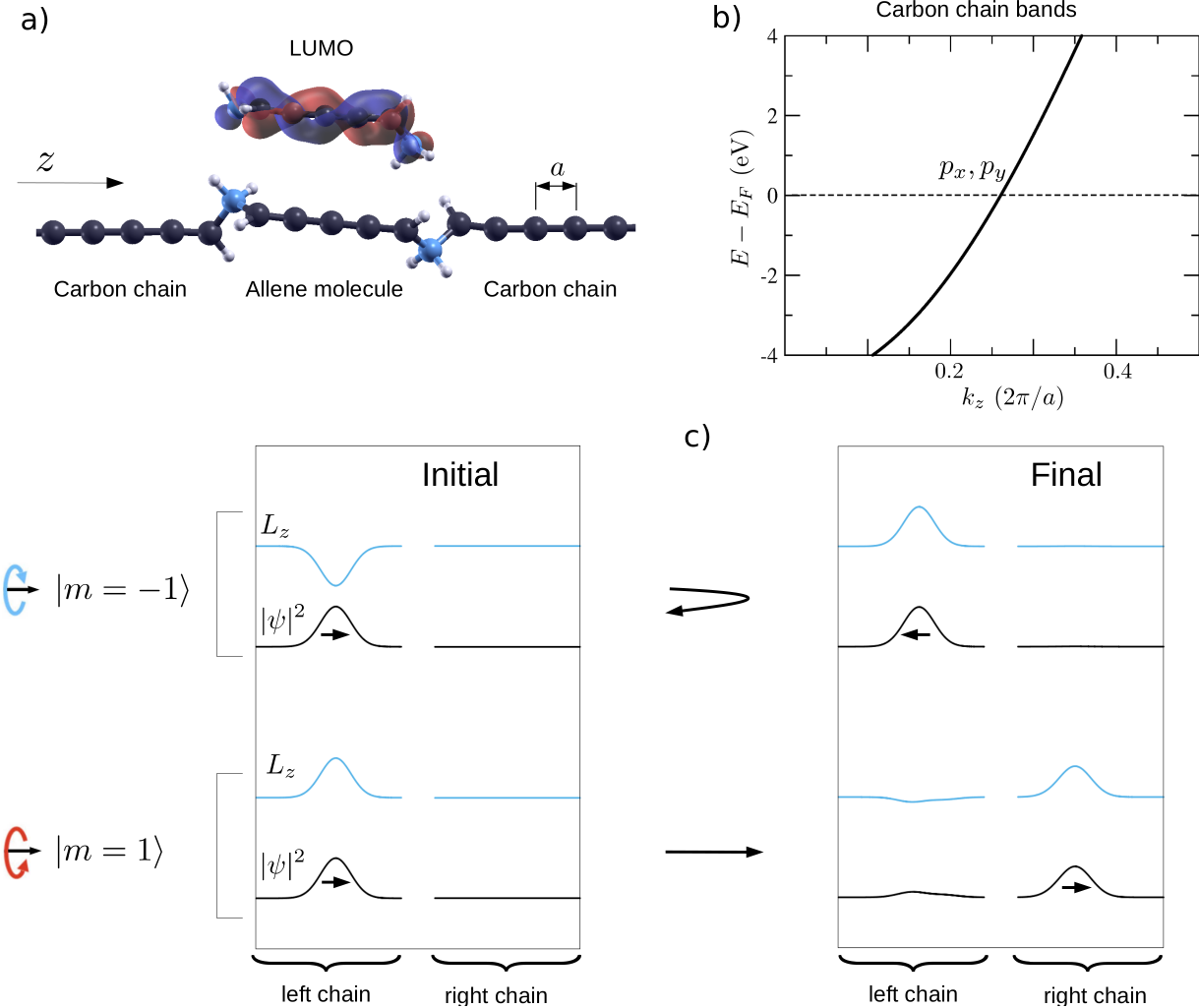}
\caption{\label{molecular_junction}
Model molecular junction showing strong orbital filtering:  
a) the junction consists of two Carbon chains connected by the 1,5-diamino-[4]cumulene (chiral Allene molecule). The shape of the LUMO is also shown;
b) the band structure of a Carbon chain with a lattice parameter of $a=1.44$~\AA; 
c) snapshots of the propagation of two wave packets, corresponding to $m=\pm 1$, across the molecular junction at initial and final times. 
}
\end{figure}

Low final orbital polarization of the wave packet in the discussed above case of the oxidized Cu surface
is related to very small orbital moments carrying by asymptotic Bloch states in the bulk copper, as can be seen in Figure~\ref{wp_junction}b, left panel. In order to have a possibility for a long range orbital current we need therefore a material where bulk asymptotic states 
allow to form a linear combination of orbitals bearing finite orbital moment.
For this we have considered a carbon chain which possesses two-fold degenerate band (per spin, we did not take into account SOC here 
which is negligible in this case) of 
$p_x,p_y$ atomic character around the Fermi energy as shown in Figure~\ref{molecular_junction}b. They can be combined to produce two propagating channels with $L_z=\pm 1$: $\left|m=\pm 1\right>=(\left|p_x\right>\pm i \left|p_y\right>)/\sqrt{2}$. 
In order to break the mirror symmetry (with respect to any plane passing through carbon chains) and mix these $p_x,p_y$ bands we connected two chains by an organic chiral molecule, 1,5-diamino-[4]cumulene\cite{MGarnerCS2019}, making a molecular junction. 
In calculations, the whole structure, shown in Figure~\ref{molecular_junction}a, was placed in a box with a vacuum of 12~{\AA}~in the $x$ and $y$ directions. 
The density of states (DOS) projected onto different molecular orbitals (see Figure~S5a in Supporting information)
shows that the LUMO (lowest unoccupied molecular orbital) is placed in the vicinity of the Fermi energy which results in a relatively large conductance as seen in Figure~S5b. This orbital (as well as other frontier orbitals shown in Figure~S5a) is chiral   
and should therefore transmit asymmetrically two $\left|m=\pm 1\right>$ channels.

Our transport calculations, shown in Figure~\ref{molecular_junction}, confirm this point.
As before, for each $\left|m=\pm 1\right>$ channel we construct in the middle of the left lead a Gaussian wave packet and let it propagate while calculating the orbital angular moment per atom along the way. The result is shown in Figure~\ref{molecular_junction}c. Here, on the left panel, we show snapshots of initial wave packet and on the right panel -- after its propagation across the molecule. 
Along with the wave packet wave function we also plot the $z$ component of the orbital moment, $L_z$. 
We can see that initially, $L_z$ is equal and opposite in sign in two channels, resulting in a net zero $L_z$. As wave packets arrive at the molecule only one of them, with $L_z=1$, is transmitted through while the other one is almost fully reflected back,
resulting in a net filtering of the orbital momentum. The resulting orbital current propagates moreover along the carbon chain 
at long distance without any decay.

\section{Conclusions}
In summary, on example of two representative systems we studied the generation and propagation of orbital moments. In the first case of Cu/O surface, where the ORE leads to the OAM texture at the Cu/O interface due to strong hybridization between Cu $d$ and O $p$ states, orbital current is found to be large only close to the interface and decays rapidly out of the oxidized surface. However, in the tunnel junction regime, the presence of the localized OAMs plays a role of a spin-dependent barrier producing a noticeable spin polarization of the current, even though SOC is quite moderate in Cu. We suggest that this spin filtering can be tuned by in-plane voltage. We may argue that similar effect could be present in other realizations where electrons traverse Oxygen layers, for example, in the case of the Oxygen intercalated inside the Cu surface; it could be probably at the origin of observed enhancement of SOT effects in oxidized Cu surfaces. 
Note that our calculations assume a phase-coherent transport regime (where the Landauer-B\"uttiker formalism is valid) while many theoretical studies on the subject have been performed within a diffusive scheme based on a kind of Boltzmann equation.
Next, in order to obtain a long range propagation of orbital current we proposed a simplistic system composed of carbon chains joined by a chiral molecule. Here, two bands of $p_x$ and $p_y$ symmetry are present in the carbon chain at the Fermi energy. They 
were shown to mix by the chiral molecule (breaking the mirror plane passing through the chains) providing two transport channels 
with a well defined (and opposite) orbital momenta, $L_z$. One of them was shown to be fully reflected while another one is transmitted which results in a long range net orbital current. Note finally that efficient orbital momentum filtering, reported here for the chiral molecule junction, may play an important role in the chirality-induced spin-selectivity (CISS) -- the effect of high interest nowadays.

\section{Methodology}
Geometric optimization and electronic properties were calculated using the Density-functional theory (DFT) as implemented in the Quantum ESPRESSO package (QE).\cite{QE1,QE2} A plane wave basis set and fully-relativistic ultrasoft pseudopodials were used for calculations. A generalized gradient approximation to exchange-correlation potential in the Perdew-Burke-Ernzerhof (PBE) parametrization was employed.\cite{PBE} An energy cut-off of 40 and 400 Ry was used for expansion of wave functions and charge densities, respectively. 
Monkhorst-Pack k-point mesh of 12x12x1 points and 1x1x4 was used for the oxidised Cu surface and the chiral molecule junction, respectively. The Hamiltonian based on maximally localized Wannier functions was then constructed using WANNIER90 code.\cite{Wann1,Wann2}.  After obtaining the Wannier Hamiltonian, we used a homemade code for wave packet propagation and subsequent orbital moment calculations using 
its matrix representation in the $s-p-d$ orbital basis. More details of calculations are given in Supporting information.

\section{Acknowledgments}
This work was supported by the French National Agency ANR programme ORION through grant no. 
ANR-20-CE30-0022-02 and by the EIC Pathfinder OPEN grant 101129641 “OBELIX”.

\end{document}


\section{Details of calculation}
\subsection{Wave packet dynamics in Wannier orbital basis}
In real space, the Wannier Hamiltonian is represented by the matrix $H_{m0, n{\bf R}}$, where $m$ and $n$ are Wannier orbitals in the reference cell 0 and the cell translated by the vector ${\bf R}$, respectively. For oxidized Cu surface we used Cu $s,p,d$ and O $s,p$ orbitals in the basis set. For the Chiral molecule we used $s,p_x,p_y$ for C, all $p$ orbitals for N and $s$ orbitals for H atoms. After obtaining the Wannier Hamiltonian for both the central (junction) region and perfect left/right leads, we used a homemade transport code TBcond implementing the non equilibrium Green’s function's approach as well as the method based on the real time propagation of electron wave packets. We give here a short description of the method, more details can be found in [1]. For wave packets, the central region representing the junction is attached on both sides to a large number, $N_L$, of lead's unit cells, simulating semi-infinite electrodes. In order to construct initial electron wave function, we first  
Fourier transform the lead's Hamiltonian: 
\begin{equation}
\label{eq:fft_lead}
H_{mn}^L({\bf k}) = \sum_{{\bf R}} \hat{H}^L_{m0, n{\bf R}} e^{i{\bf kR}}.
\end{equation}
and then diagonalize it: 
$$
\sum_{n} H_{mn'}^L({\bf k}) u_{n'n}({\bf k}) = \varepsilon_n({\bf k}) u_{mn}({\bf k}). 
$$
This provides us with electronic bands $\varepsilon_n({\bf k})$ the number of which is determined by the number of Wannier functions
in the lead unit cell. Corresponding Bloch states at the orbital $m$ in the cell ${\bf R}$ of the whole lead are given by Bloch theorem:
\begin{equation}
\label{eq:bloch_function}
\varphi_{n{\bf k}} (m,{\bf R}) = u_{mn}({\bf k}) e^{i{\bf kR} }.
\end{equation}
The band structure is used to select right-moving Bloch states at the energy of interest (usually the Fermi energy), 
determining possible {\bf k} and $n$ as different transport channels. 
The initial wave packet for each channel is constructed by weighting the Bloch state by the Gaussian function:
\begin{equation}
\label{eq:wave_function_wp2}
\varphi({m,\bf R}) = A e^{- \frac{ \left|{\bf R-R}_0\right|^2} {\gamma^2} }\cdot  \varphi_{n{\bf k}} (m,{\bf R}).
\end{equation}
The parameter $\gamma$ defines the spatial localization of the wave packet, while ${\bf R}_0$ -- its starting unit cell
which is normally chosen to be in the middle of the left electrode, at $N_L/2$. $A$ is the normalization constant.  
The wave packet is then evolved in time with the Hamiltonian matrix using Schr\"odinger equation  ($i\hbar\frac{\partial \psi}{\partial t}=H\psi$) by applying an approach based on Chebyshev polynomials [2,3].

The propagation time can be specified either explicitly or estimated from the total length of the system (in the $z$ direction) and 
the group velocity of the wave packet. The latter is calculated using the velocity operator defined as following :
\begin{equation}
\label{eq:velocity_wp}
\hat{{\bf V}} = \frac{d \hat{\bf r} (t)}{dt} = -\frac{i}{\hbar} [\hat{\bf r}, \hat{H}^L]
\end{equation} 
Applied to the Bloch state, $\varphi_{n{\bf k}}$, it produces:   
\begin{equation}
{\bf V}_{n{\bf k}} = -{i \over \hbar} \sum_{mm'{\bf R}'} \varphi_{n{\bf k}}(m,0)^* \left[ {\bf r}_{m}\cdot H^L_{m0,m'{\bf R}'} - 
H^L_{m0,m'{\bf R}'}\cdot ({\bf R}'+{\bf r}_{m'}) \right] \varphi_{n{\bf k}}(m',{\bf R}'),
\end{equation}
where $m,m'$ run, as usual, over Wannier functions in the unit cell, ${\bf R}$ goes over all translation vectors, and
${\bf r}_{m}$ stay for position vectors of the orbital $m$ within the unit cell.

We used $N_L=160$ and 480 for the oxidized Cu surface and the chiral molecule, respectively, which 
translates into the corresponding propagation time of 132 and 57 fs.

\subsection{Orbital moments in Wannier basis}
The orbital moment's calculation for $p$ states was done using the following matrix representation of $\textbf{L}^{p}=(L_x^{p}, L_y^{p}, L_z^{p})$, with basis set as $|p_z\rangle$, $|p_x\rangle$, $|p_y\rangle$

\begin{equation}
L_x^{(p)} = \hbar \begin{pmatrix}
 0 & 0 & i \\
 0 & 0 & 0 \\
-i & 0 & 0
\end{pmatrix}
\end{equation}

\begin{equation}
L_y^{(p)} = \hbar \begin{pmatrix}
 0 & -i & 0 \\
 i &  0 & 0 \\
 0 &  0 & 0
\end{pmatrix}
\end{equation}

\begin{equation}
L_z^{(p)} = \hbar \begin{pmatrix}
 0 &  0 &  0 \\
 0 &  0 & -i \\
 0 &  i &  0
\end{pmatrix}
\end{equation}

Similarly, the matrix representation for $\textbf{L}^{d}=(L_x^{d}, L_y^{d}, L_z^{d})$, with basis set as $|d_z^2\rangle$, $|d_{xz}\rangle$, $|d_{yz}\rangle$, $|d_{x^2-y^2} \rangle$ and $|d_{xy}\rangle$ is as follows:

\begin{equation}
L_x^{(d)} = \hbar \begin{pmatrix}
 0 & 0 & \sqrt{3}i & 0 & 0 \\
 0 & 0 & 0 & 0 & i \\
 -\sqrt{3}i & 0 & 0 & -i & 0 \\
 0 & 0 & i & 0 & 0 \\
 0 & -i & 0 & 0 & 0 
\end{pmatrix}
\end{equation}

\begin{equation}
L_y^{(d)} = \hbar \begin{pmatrix}
 0 & -\sqrt{3}i & 0 & 0 & 0 \\
 \sqrt{3}i & 0 & 0 & -i & 0 \\
 0 & 0 & 0 & 0 & -i \\
 0 & i & 0 & 0 & 0 \\
 0 & 0 & i & 0 & 0 
\end{pmatrix}
\end{equation}

\begin{equation}
L_z^{(d)} = \hbar \begin{pmatrix}
 0 & 0 & 0 & 0 & 0 \\
 0 & 0 & -i & 0 & 0 \\
 0 & i & 0 & 0 & 0 \\
 0 & 0 & 0 & 0 & -2i \\
 0 & 0 & 0 & 2i & 0 
\end{pmatrix}
\end{equation}

The spin moments were calculated using the Pauli matrices:
\begin{equation}
\hat{S}_x = \frac{\hbar}{2} \begin{pmatrix}
0 & 1 \\
1 & 0
\end{pmatrix}
\end{equation}

\begin{equation}
\hat{S}_y = \frac{\hbar}{2} \begin{pmatrix}
0 & -i \\
i & 0
\end{pmatrix}
\end{equation}

\begin{equation}
\hat{S}_z = \frac{\hbar}{2} \begin{pmatrix}
1 & 0 \\
0 & -1
\end{pmatrix}
\end{equation}

OAM expectation value at the Fermi surface, shown in Figure~1d, was calculated as:
\begin{equation}
\textbf{L}_{FS}(\textbf{k})=-4k_BT\sum_nf^{'}_{n\textbf{k}}\langle u_{n\textbf{k}}|\textbf{L}|u_{n\textbf{k}}\rangle \\
                           =\sum_n \frac{2\langle u_{n\textbf{k}}|\textbf{L}|u_{n\textbf{k}}\rangle}{1+\text{cosh}[(E_{n\textbf{k}}-E_F)/k_BT]},
\end{equation}
where $k_B$ is the Boltzmann's constant, $T$ is the Temperature, 
$|u_{n\textbf{k}}\rangle$ is the periodic part of the Bloch state with $n$ as the band index, and $f^{'}_{n\textbf{k}}$ is the energy derivative of the Fermi-Dirac distribution function. $k_BT$= 25 meV has been used in the paper.

The Kubo formula used to calculate the orbital moment accumulation in Figure~1b of the paper is as follows:
\begin{equation}
\delta L_x =-e\mathcal{E}_y\tau\frac{1}{N_{\textbf{k}}}\sum_{n\textbf{k}}f^{'}_{n\textbf{k}} \langle u_{n\textbf{k}}|L_x|u_{n\textbf{k}}\rangle \langle u_{n\textbf{k}}|v_y|u_{n\textbf{k}}\rangle,
\end{equation}
where $e$ is the magnitude of electron's charge, $\tau$ is the average electron momentum relaxation time, $\mathcal{E}_y$ is the strength of electric field applied along the $y$ direction, $v_y$ is the $y$ component of the velocity operator, and $N_{\textbf{k}}$ is the number of $\textbf{k}$ points used for the integral. In our calculations we have used a mesh of 260x260 \textbf{k}-points.

We note that the quantity $e\mathcal{E}_y\tau/\hbar$ can be seen as the amount of the Fermi surface displacement in the \textbf{k}-space due to applied electric field, $\delta k_y=e\mathcal{E}_y\tau/\hbar$. Therefore, in Figure~1b we have chosen to plot the quantity:
\begin{equation}
\delta L_x /\delta k_y =-\frac{\hbar}{N_{\textbf{k}}}\sum_{n\textbf{k}}f^{'}_{n\textbf{k}} \langle u_{n\textbf{k}}|L_x|u_{n\textbf{k}}\rangle \langle u_{n\textbf{k}}|v_y|u_{n\textbf{k}}\rangle
\end{equation}
which describes an OAM response to applied electric field and has dimensionality of [$\hbar\cdot$\AA].

\section{Bandstructure of the Cu Lead}
\begin{figure}[H]
 \includegraphics[width=17cm]{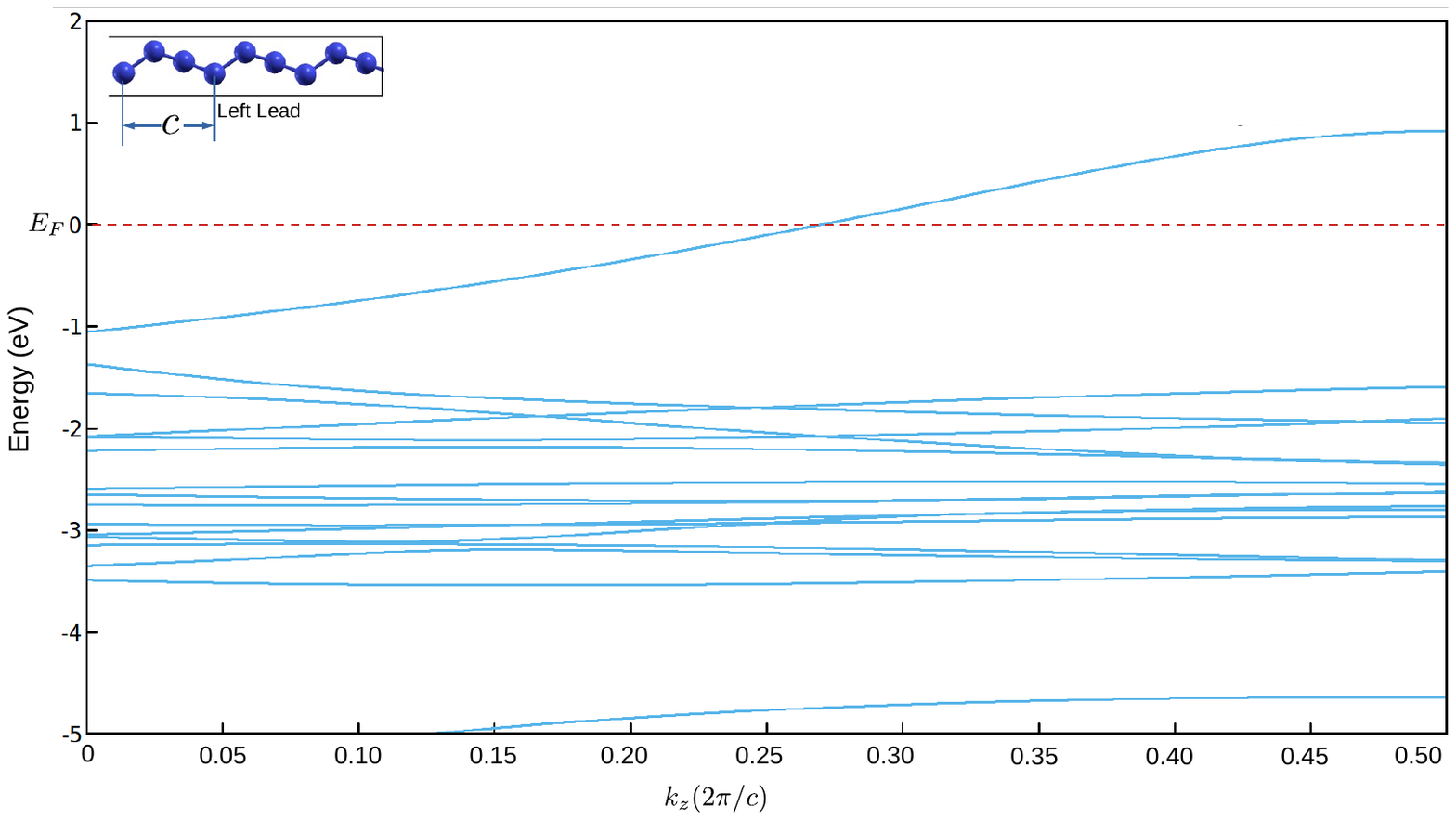}
  \caption{Bandstructure of Cu lead at $k_{||}=0.15\Vec{b_2}$ of the Brilloun zone, between $k_z=0.00\Vec{c}$ and $k_z=0.50\Vec{c}$. }
  \label{FigS1}
\end{figure}

Bandstructure of Cu lead (bulk configuration) at $k_{||}=0.15\Vec{b_2}$ of the Brilloun zone, between $k_z=0.00\Vec{c}$ and $k_z=0.50\Vec{c}$ was calculated and is shown in Figure \ref{FigS1}. This shows the availability of energy eigenvalues at Fermi energy.

\section{Oxidized Cu surface without Spin-Orbit Coupling}
\begin{figure}[H]
 \includegraphics[width=17cm]{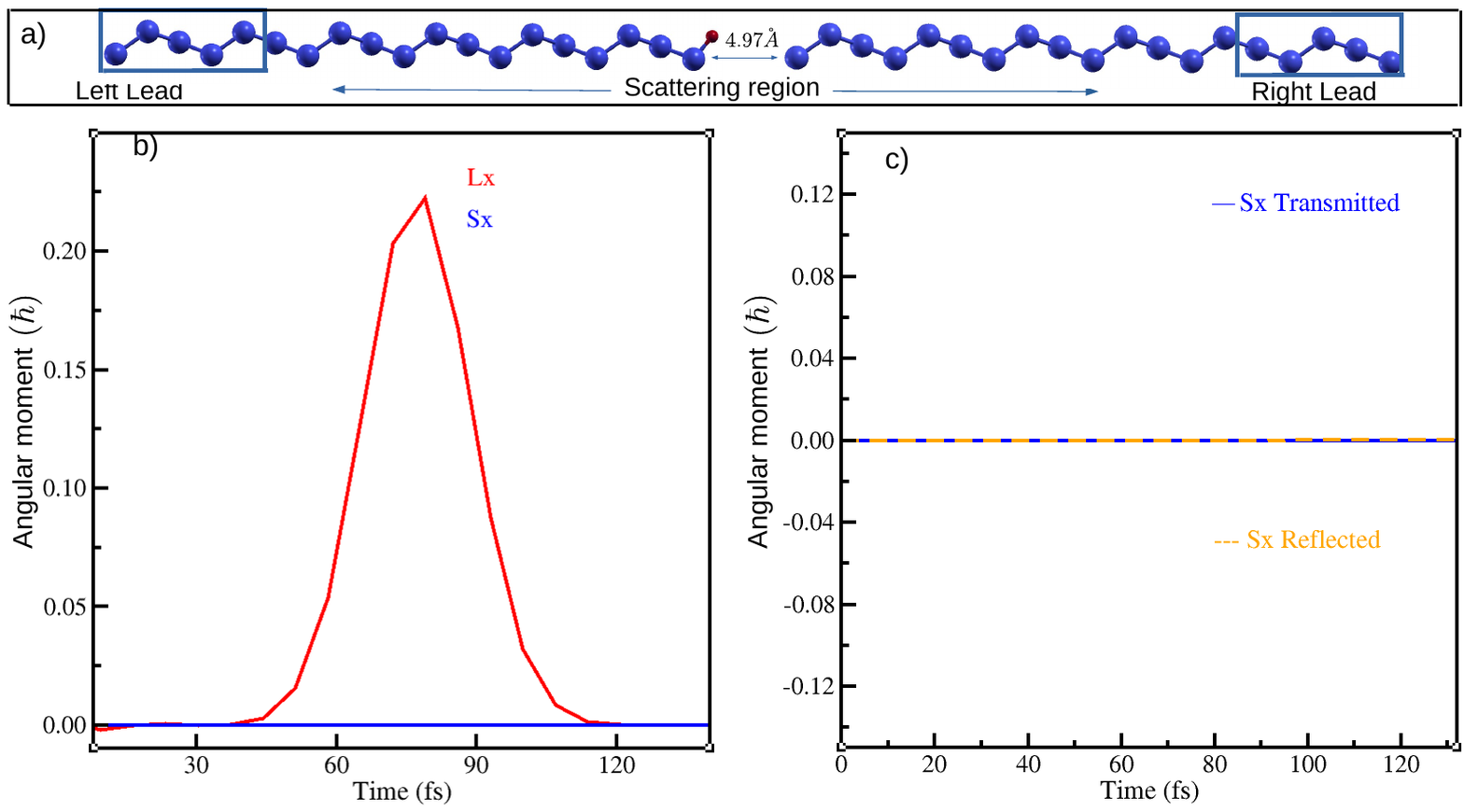}
  \caption{a) Oxidized Cu surface with leads attached on both ends, here surface Cu-Cu distance is 4.90 \AA. b) Total $L_x$ (red) and $S_x$ (blue) summed over both the propagating channels. c) $S_x$ for transmitted (blue) and reflected (orange) part of the wavepacket. }
  \label{FigS2}
\end{figure}

Oxidized Cu surface, where the distance between the two Cu surface atoms is 4.97 \AA, with leads attached on both sides, is shown in Figure \ref{FigS2}a. In this configuration, the caclulation was done without considering spin-orbit coupling interaction. In this case we propagated the Gaussian wavepacket from the middle of the left lead to the middle of the right lead and calculated the total $L_x$ and total $S_x$ summed over both the channels. These are shown in Figure \ref{FigS2}b. We also calculated the $S_x$ for the transmitted and reflected part of the wavepacket. This is shown in Figure \ref{FigS2}c. A finite $L_x$ was observed in this case even without spin-orbit coupling (SOC), indicating that SOC is not needed for generation of orbital moment. However due to the lack of SOC, there is no spin current.

\section{Cu Surface}
\begin{figure}[H]
 \includegraphics[width=17cm]{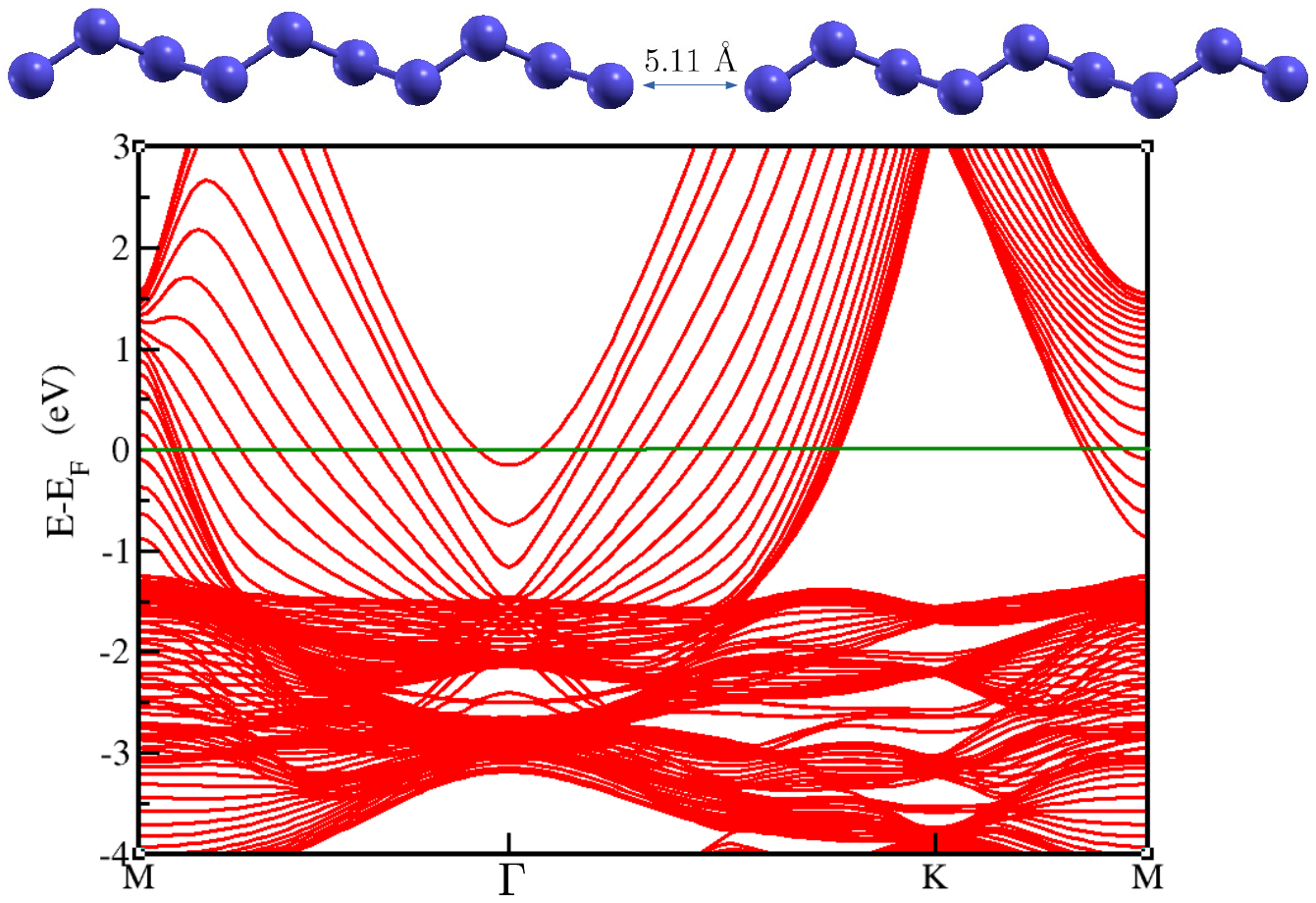}
  \caption{a) Cu surface with the distance between the two Cu surface atoms 5.11 \AA. b) Bandstructure of this structure along various symmetry directions of the Brillion zone  }
  \label{FigS3}
\end{figure}

\begin{figure}[H]
 \includegraphics[width=17cm]{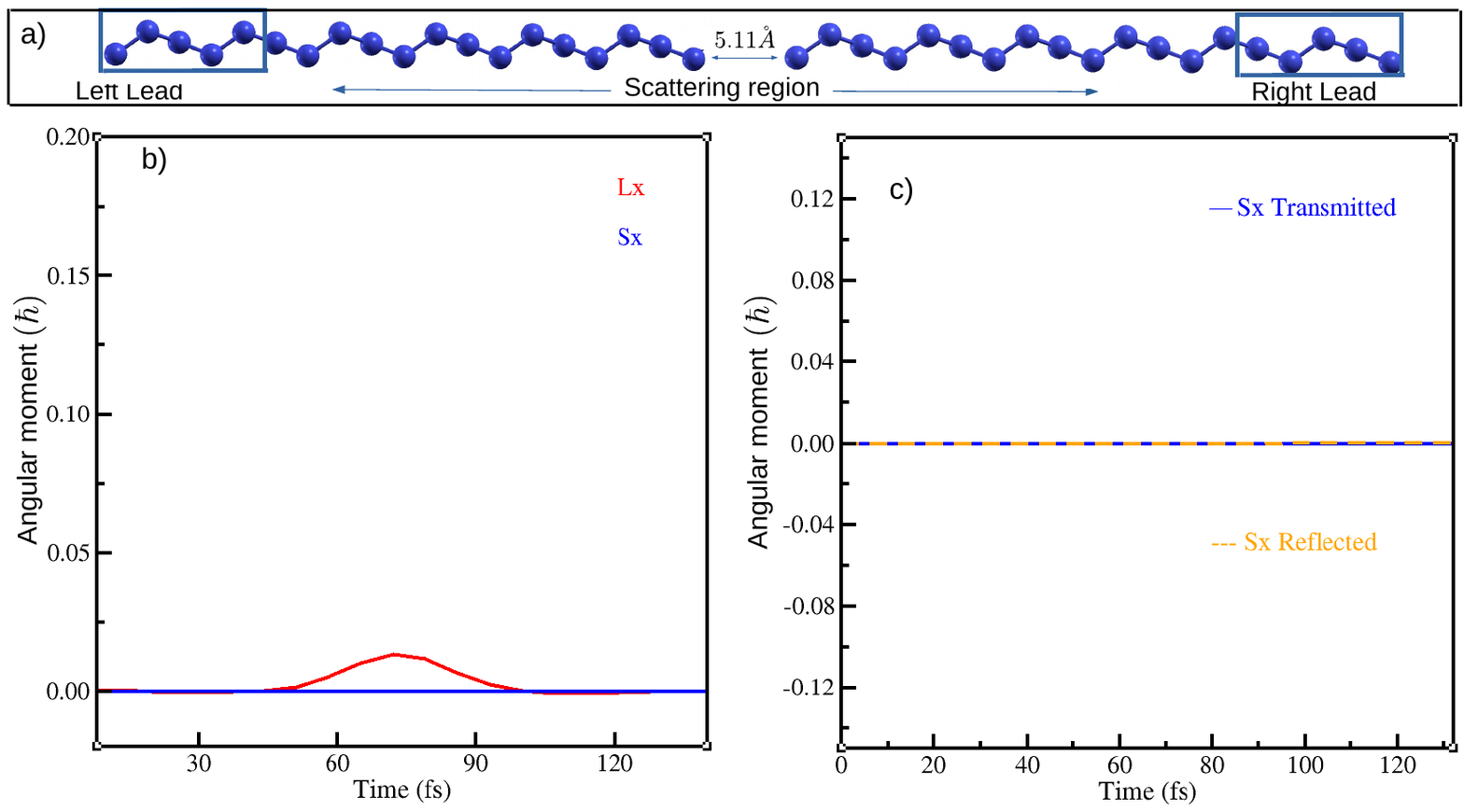}
  \caption{a) Cu surface with leads attached on both ends. b) Total $L_x$ (red) and $S_x$ (blue) summed over both the propagating channels. c) $S_x$ for transmitted (blue) and reflected (orange) part of the wavepacket. }
  \label{FigS4}
\end{figure}

Cu surface with the distance between the two Cu surface atoms 5.11 \AA, as shown in Figure \ref{FigS3}a. For this structure the bandstructure along various symmetry directions of the Brillion zone was calculated and is shown in Figure \ref{FigS3}b. The same structure with Cu leads attached on both sides is shown in Figure \ref{FigS4}a. For this structure also we propagated a Gaussain wavepacket from the middle of the left lead to the middle of the right lead and calculated total $L_x$ and $S_x$ summed over the two channels. This is shown in Figure \ref{FigS4}b. Similar to the other cases here also we calculated $S_x$ for the transmitted and reflected part of the wavepacket. This is shown in Figure \ref{FigS4}c. These calculations were done by considering SOC interaction. In this structure, as there is no O, the $d$ orbitals of the surface Cu atom alone doesn't result in generation of significant orbital moment, and subsequently no generation of spin current. 

\section{Chiral molecule}
\begin{figure}[H]
 \includegraphics[width=17cm]{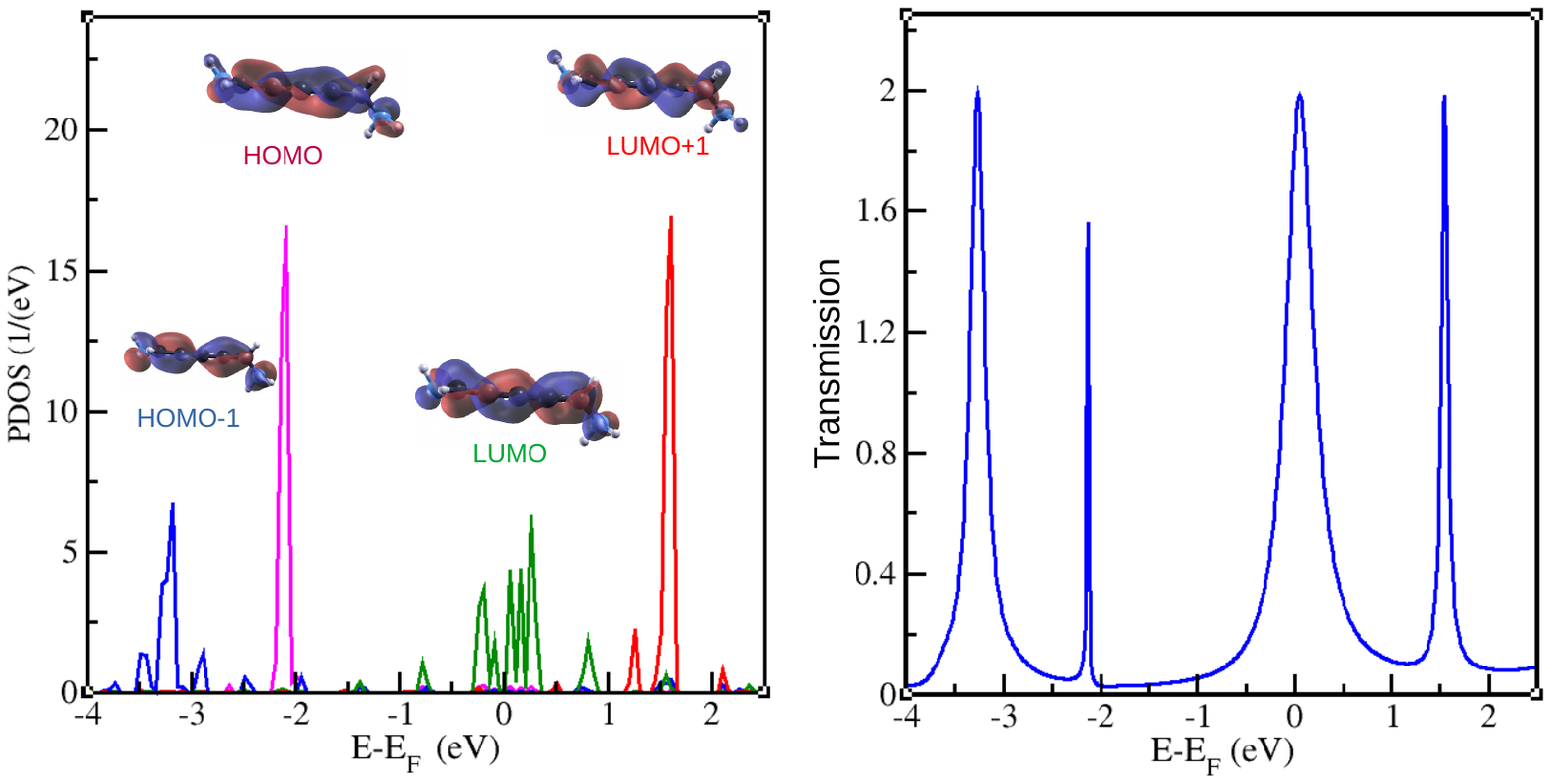}
  \caption{a) The density of state (DOS) projected on to the isolated molecule, for highest occupied molecular orbital (HOMO), HOMO-1, lowest unoccupied molecular orbital (LUMO) and LUMO-1. The orbital characters of HOMO, HOMO-1, LUMO and LUMO-1 is also shown here. b) The transmission profile of the carbon chain with the chiral molecule attached to it}
  \label{FigS5}
\end{figure}

For the system shown in Figure 5a, the density of state (DOS) projected on to the isolated molecule, for highest occupied molecular orbital (HOMO), HOMO-1, lowest unoccupied molecular orbital (LUMO) and LUMO-1 is shown in Figure \ref{FigS5}a. The orbital characters of HOMO, HOMO-1, LUMO and LUMO-1 is also shown in this figure. The transmission profile for this structure is shown in Figure \ref{FigS5}b.

